\documentclass[conference]{IEEEtran}        
\usepackage{amsmath}
\usepackage{amsfonts}
\usepackage{amssymb}
\usepackage{graphicx}
\usepackage{float}
\usepackage{algorithmicx}

\floatstyle{ruled}
\newfloat{algorithm}{tbp}{loa}
\providecommand{\algorithmname}{Algorithm}
\floatname{algorithm}{\protect\algorithmname}

\title{\LARGE \bf
Maximizing Profit of Cloud Brokers under Quantized Billing Cycles: a Dynamic Pricing Strategy based on Ski-Rental Problem}

\author{Gourav Saha and Ramkrishna Pasumarthy%
\thanks{Gourav Saha is a graduate student in the Systems and Control Group, Department of Electrical Engineering, IIT Madras, India.
        {\tt\small ee13s005@ee.iitm.ac.in}}%
\thanks { Ramkrishna Pasumarthy is with the Faculty of Electrical Engineering, IIT Madras, India.
 {\tt\small ramkrishna@ee.iitm.ac.in}}
 }
\begin{document}

\maketitle
\thispagestyle{empty}
\pagestyle{empty}

\begin{abstract}
In cloud computing, users scale their resources (computational) based on their need. There is massive literature dealing with such resource scaling algorithms. These works ignore a fundamental constrain imposed by all Cloud Service Providers (CSP), i.e. one has to pay for a fixed minimum duration irrespective of their usage. Such quantization in billing cycles poses problem for users with sporadic workload. In recent literature, Cloud Broker (CB) has been introduced for the benefit of such users. A CB rents resources from CSP and in turn provides service to users to generate profit. Contract between CB and user is that of pay-what-you-use/pay-per-use. However CB faces the challenge of Quantized Billing Cycles as it negotiates with CSP. We design two algorithms, one fully online and the other partially online, which maximizes the profit of the CB. The key idea is to regulate users demand using dynamic pricing. Our algorithm is inspired by the Ski-Rental problem. We derive competitive ratio of these algorithms and also conduct simulations using real world traces to prove the efficiency of our algorithm.
\end{abstract}

\section{INTRODUCTION}
\label{ch:chap8}

\subsection{Overview}
\label{overview}

There is no universal definition of cloud computing. However as far as our research is concerned, the most apt definition of cloud computing is found in \cite{berkeley_view} which can be quoted as: "computing as a \textit{utility}". In our day to day life the most common utilities are electricity, water, gas, heat, postpaid mobile services etc.   Similarly in cloud computing, computing resources (like CPU, memory, storage, network domains, virtual desktop) are \textit{rented} to users based on their demand. From user's viewpoint, it eliminates the need of an upfront investment as an user can pay based on the amount of resources  it has used. This is termed as ``pay-per-use'' or ``pay-as-you-go'' model. Therefore, \textit{resource scaling} is the most fundamental aspect of cloud computing and hence extensive amount of effort has been channeled to explore this area. Cloud Service Providers (or CSP's), like Amazon, ElasticHost etc, rent computing resource to the users in form of \textit{Virtual Machines} (also called \textit{instances}) or VMs. Scaling of VMs revolves around two fundamental questions:

\begin{itemize}

\item[\textbf{1.}] \textit{From Users Perspective: }How to scale VMs to optimize a certain objective?

\item[\textbf{2.}] \textit{From CSP's Perspective: }How to support the active VMs using minimum number of physical servers?

\end{itemize}

In cloud computing literature the former is often called \textit{auto scaling} while the latter is called \textit{dynamic provisioning}. These two questions are indeed similar and rely on a common line of research. Various researchers approached these two problems with different objectives and by using different mathematical tools. In the following we will give a brief account of these approaches.

Concerning auto scaling, the most simplest methods are the Static Threshold based policies \cite{staticthreshold} where the scaling of VMs is triggered when certain CPU parameters (CPU load, response time) crosses a pre-defined threshold. Such methods are too simple and if not designed properly face the problem of limit-cycles. Queuing theory based techniques are also very well developed. Some well cited papers in this regard can be \cite{queue2,queue3}. Theoretic results available in queuing theory uses simplified models and hence cannot be directly applied to complex systems like those encountered in cloud computing framework. Also the results are statistical implying that the analysis is true only over a long run after the system reaches steady-state. Control Theory based approaches aim to solve the latter problem by trying to control the transients \cite{control3,control4}. They are generally reactive methods, and hence are not good for workloads which have a lot of sudden spikes. There is a lot of active research concerning proactive methods. They depend on predictive abilities and use tools like Model Predictive Controller \cite{mpc}, Time Series Analysis \cite{timeseries1,timeseries2}, Machine Learning \cite{PhDThesisML}. Work done in \cite{timeseries2} is impressive in the sense that the prediction algorithm has low overhead and also has the ability to distinguish fine workload patterns. There are several tasks which are deadline sensitive. There is a separate line of research based on heuristic algorithms \cite{deadline1,deadline2} to handle these tasks.

Similar tools have been used to optimize performance of data centers. Queuing theory has been widely used. \cite{queue1} is a good reference in this regard as it uses much less approximation to model a cloud data center. Control Theoretic approaches are very famous. One popular book dealing with this subject is \cite{control1}. \cite{control2} is as well cited paper which presents practically viable method to regulate a performance metric of data center using feedback control. A very interesting method to save energy of a cloud data center has been presented in \cite{control5} where they use feedback control to regulate CPU clock frequency based on incoming request rate. Optimization based techniques have been directly used to design resource allocation strategies to maximize profit of cloud data centers \cite{optimization1,optimization2}. Another line of research considers optimizing the electricity and the cooling cost required to run a data center. Such problems are not trivial when electricity \cite{electricity} and the water cost \cite{cooling} are time varying.

\begin{figure}[t]
\begin{centering}
\includegraphics[scale=0.65]{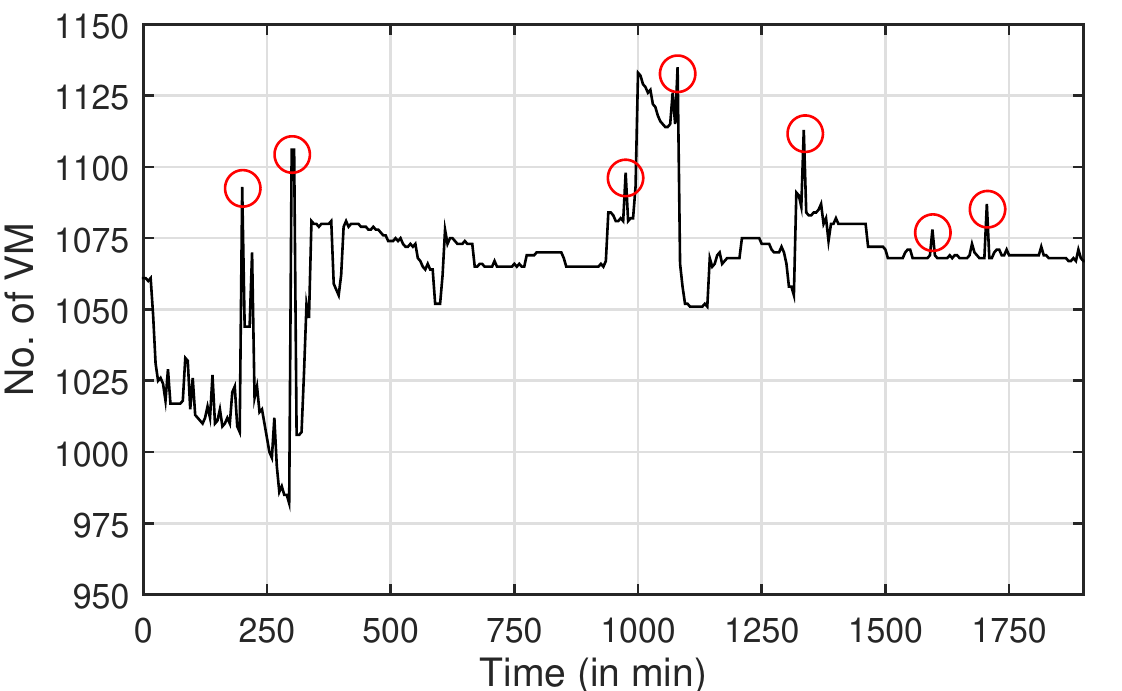}
\par\end{centering}

\caption{A part of the google cluster usage traces showing spikes in user demand. Red circles indicates the spikes.}

\label{fig:spiky_graph}
\vspace{-1.0em}
\end{figure}

\subsection{Quantized Billing Cycles and the Cloud Broker}
\label{QBC}

There is a fundamental misconception regarding resource scaling in cloud computing literature. Definitely, the main objective of cloud computing is elasticity, i.e. the user can scale-up or scale-down VMs based their short term needs. However this ``short term'' is not infinitely small in the sense that it may not be possible to return the VMs in the very next instant after it was rented.

We will explain this situation using an example. Consider that we are using on-demand instances of Amazon EC2 to satisfy these demands. Billing Cycle of AmazonEC2 on demand instance is $1$ hour, i.e. you have to pay the same price if you use the VM for $1$ min or $1$ hour. We call this phenomenon as \textit{Quantized Billing Cycles} (abbreviated as QBC) in the rest of the paper. This leads to serious problem especially for those users whose demand pattern is sporadic in nature. For e.g. Consider the workload trace shown in Figure \ref{fig:spiky_graph} which shows the number of VMs required to satisfy user demand. There are few spikes in the demand curve to satisfy which one needs to buy extra VMs. These VMs may be in use for a very small fraction of $1$ hour after which it may be idle. However we cannot return these VMs before $1$ hour\footnote{The user may choose to return the VM but it won't get any financial benefit. Therefore it is better to keep the VM for the entire billing cycle even though it remains idle. This is called \textit{smart-kill}.}. Hence there will be wastage of VMs.

\begin{figure}[t]
\begin{centering}
\includegraphics[scale=0.45]{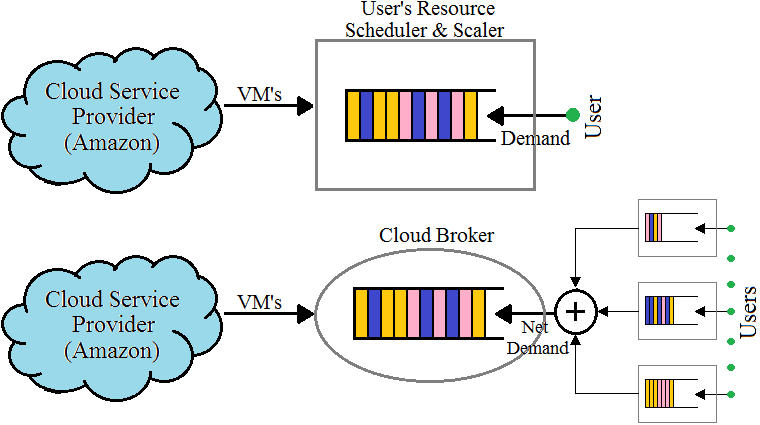}
\par\end{centering}

\caption{Schematic showing cloud broker as a middleman between Cloud Service Provider (CSP) and the users. Adapted from \cite{main2}.}

\label{fig:broker}
\vspace{-0.7em}
\end{figure}

To mitigate this problem to some extent the concept of cloud broker has been proposed in recent literature \cite{broker,broker_lit,main2}. A broker forms a middle man between the CSP and the general users as shown in Figure \ref{fig:broker}. The users send their job requests to the cloud broker. The cloud broker rents VMs from the Cloud Service Provider to service these demands. The cloud broker charges the user based on the fraction of VMs resources used to service the job request. This is called \textit{pay-what-you-use}. It can also be based on \textit{per-request-basis}. This transfers the challenge of QBC from the users to the cloud broker. However this issue is not as critical for the cloud broker as it is for the users. This can be understood as follows:

\begin{itemize}

\item[\textbf{1.}] QBC poses problems for those users whose demand is sporadic. Higher the sporadic nature, greater the loss.

\item[\textbf{2.}] Cloud broker serves the aggregate demand of many users. In statistical sense, the sporadic/spiky nature in aggregated demand should be lesser compared to individual user's demand. This is because a summer is a discrete integrator, a low pass filter which removes the noises.

\end{itemize}

In the rest of the paper we will concentrate on how a cloud broker can maximize its profit under QBC.

\textit{Remark: }We would like to stress that the mathematical formulation, algorithms and analysis presented here can be applied to a single user without any change. However given that a cloud broker has a smoother workload pattern compared to a single user, a broker will benefit more compared to a user.

\subsection{Existing literature and our contribution}
\label{sec:contribution}

In our problem, every time a cloud broker has to buy a VM it faces the risk of under utilization of the VM in the subsequent time slots. The broker has to make a decision without knowledge of future demand. Study of such problems comes under the category of Online Algorithms, more specifically the Ski-Rental problems. There have been a few applications of Ski-Rental literature to solve real world problems. In this section we will discuss the ones pertaining to cloud computing. These works has close resemblance with our problem.

We derive our motivation from the work done in \cite{main1,main2}. In these two works the cloud broker has to decide in each time slot whether to reserve instances or to serve the demand using on-demand instances. If the demand persists for a long time then reserving is a better option. However if the demand falls down quickly it is better to use on-demand instances. Cloud broker has to make this decision without any knowledge of future demand or with partial knowledge of future demand. The authors designed both deterministic and randomized online algorithms and also derived their competitive ratio. \cite{main4} and \cite{MainFuture} considers the same problem however the model used in \cite{main4} is more general. It tries to optimize the switching cost and electricity cost of a data center. If there is a decrease in demand then the data center has to decide whether to shut down few physical servers or let them run. To run a physical server one needs to bear the electricity bill while switching down the server incurs wear-and-tear cost. Wear-and-tear cost is higher than electricity bill in short run but smaller in the long run. If one decides to switch off the server and immediately later it is needed, then one saves a small electricity bill in the expense of suffering a large switching cost. Therefore taking such decisions without knowledge of future demand is difficult. In \cite{main4}, author designed only deterministic algorithms to tackle this problem while \cite{MainFuture} considers both deterministic and randomized algorithms. A recent work presented in \cite{constrained_skirental} shows that the competitive ratio of the randomized algorithms can be improved if statistical information (like first and second order moment) of the demand process is available.

To the best of our knowledge the references given above constitutes almost all the major work dealing with the use of ski-rental framework for cloud computing applications. We make two main contributions to this literature:

\begin{itemize}

\item[\textbf{1.}]We suggest how \textit{dynamic pricing} can be used to maximize the profit of the cloud broker under QBC. This can be used as an alternative to the work done in \cite{main1,main2} specially in the case where the workload is deferrable. One can also use this approach alongside \cite{main1,main2}. These two points will be discussed further in Section \ref{ch:chap11}.

\item[\textbf{2.}]We show that the knowledge of future demand leads to better competitive ratio for the Partial Online algorithm. In \cite{MainFuture} such attempts have been made but with a major assumption on the demand graph, i.e. the demand increases or decreases no more than one step in every time slot.

\end{itemize}

\section{Problem Formulation}
\label{ch:chap9}

\subsection{Motivating Example}

We aim to use dynamic pricing as a control signal to regulate user demand. The motivation to use dynamic pricing in cloud computing setup is derived from the paper \cite{main3}. Before proceeding forward with the quantitative formulation of the problem, we will first consider an example to illustrate our idea. Consider the following scenario:

\begin{itemize}

\item[\textbf{1.}]Cloud broker buys VMs from Amazon at $\$0.132$ per VM. This is the cost price of the cloud broker. The billing cycle of each VM is $1$ hour $= 60$ min.

\item[\textbf{2.}]Duration of a time slot = $10\, min$. Hence a VM is active for $6$ time slots.

\item[\textbf{3.}]Nominal selling rate = $\$0.03$ per VM per time slot. Nominal selling rate should be such that if the VM is in use for all the time it is active ($6$ time slots here) then the cloud broker should make a profit. In our case: $\$0.03 \times 6 = \$0.18 > \$0.132$. Hence the condition holds.

\item[\textbf{4.}]The cloud broker has the freedom of changing the selling rate\footnote{In the work the term ``VM price'' inherently means the selling price of the VM not the cost price.} at the beginning of each $10\; min$ interval. The users remains totally aware of the current selling rate.

\end{itemize}

\begin{figure}[t]
\begin{centering}
\includegraphics[scale=0.43]{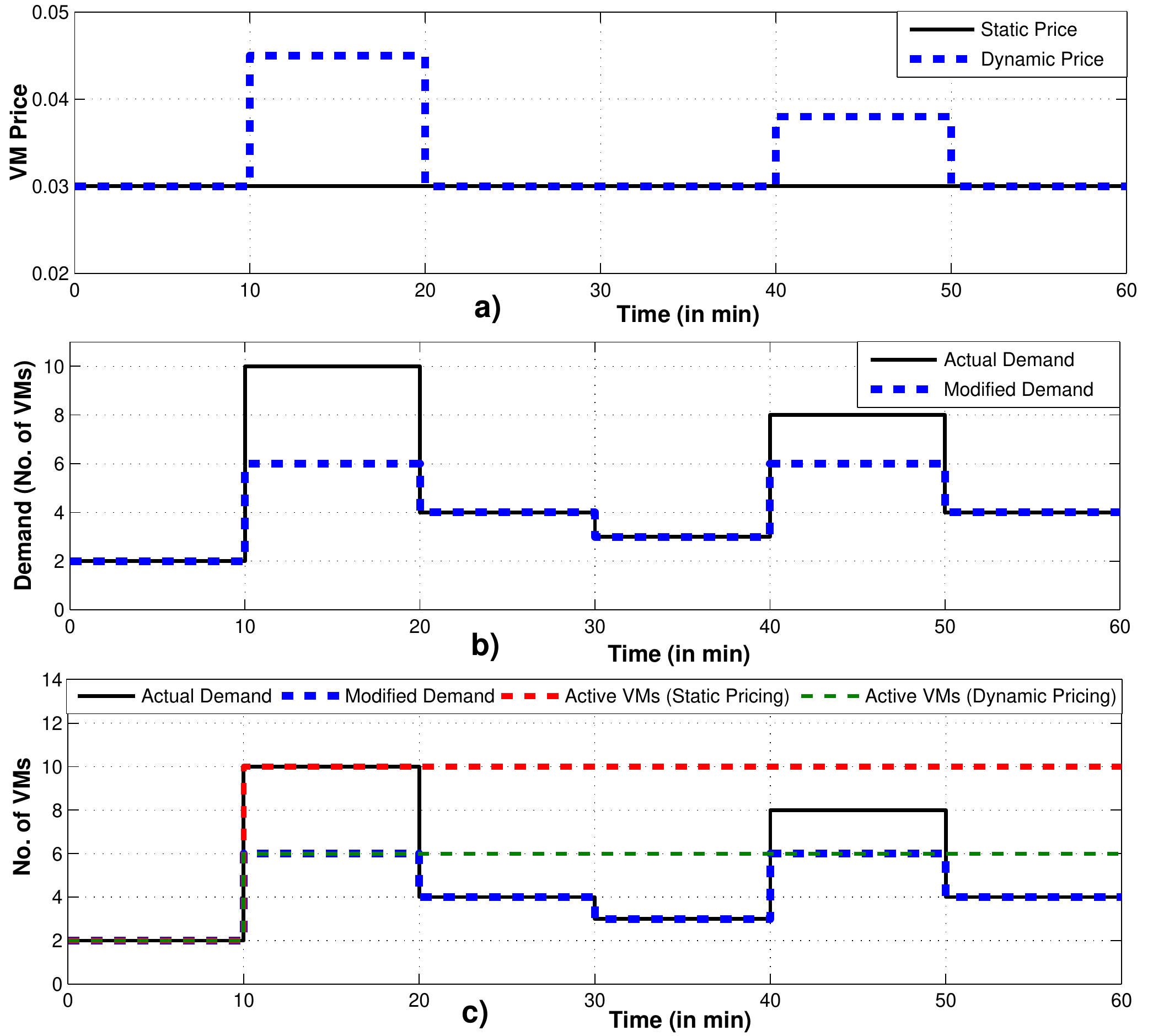}
\par\end{centering}

\caption{An example of static and dynamic pricing: a) Graph showing Price vs Time for static and dynamic pricing. b) The graph corresponding to actual demand (for static pricing) and modified demand (for dynamic pricing). c) Graph showing actual/modified demand along with the active number of VMs for both static and dynamic pricing case.}

\label{fig:motivating_example_graph}
\vspace{-0.7em}
\end{figure}

We will investigate two cases, one with static pricing and the other with dynamic pricing. Relevant graphs are shown in Figure \ref{fig:motivating_example_graph}. In both the cases we start with the assumption that at time $t=0$ the cloud broker has no VMs.

\noindent \textit{Case 1 (Static Pricing):}

In static pricing, the selling price remains constant at nominal rate therefore, $\text{{Demand}}=\text{{Actual}}\:\text{{Demand}}$.
In the $1^{st}$ interval the demand is of $2$ VMs while we have $0$
VMs. So we buy $2$ VMs incurring a cost of $2\times\$0.132=\$0.264$.
In the $2^{nd}$ interval, demand is of $10$ VMs while we have $2$
VMs (bought in the $1^{st}$ interval). So we buy $8$ VM incurring
a cost of $8\times\$0.132=\$1.056$. In the next $4$ intervals the
demand is less than $10$ VM and hence we don't have to buy any more
VMs. Therefore the net cost price of the cloud broker is $\$1.32$.
The net selling price of the cloud broker is $\left(2+10+4+3+8+4\right)\times\$0.03=\$0.93$.
$\text{{Profit}}=\text{{Selling}}\:\text{{Price}}-\text{{Cost}}\:\text{{Price}}=\$0.93-\$1.32=-\$0.39$,
i.e. the cloud broker suffered a loss.

\noindent \textit{Case 2 (Dynamic Pricing):}

In this case the selling price varies in response to which the user's
demand gets modified. This case is similar to \textit{Case 1} except
that in the $2^{nd}$ and the $5^{th}$ interval the selling price
goes up to $\$0.045$ and $\$0.038$ respectively. The cloud broker
has to buy $2$ VMs and $4$ VMs in the $1^{st}$ interval and $2^{nd}$
interval respectively incurring a cost price of $6\times\$0.132=\$0.792$.
The net selling price of the cloud broker is $\left(2+4+3+4\right)\times\$0.03+6\times\$0.045+6\times0.038=\$0.888$.
$\text{{Profit}}=\text{{Selling}}\:\text{{Price}}-\text{{Cost}}\:\text{{Price}}=\$0.888-\$0.792=\$0.096$,
i.e. the cloud broker makes a profit.

We will encapsulate the idea behind dynamic pricing by making the following key observations:

\begin{itemize}

\item[\textbf{1.}]The idea of increasing the selling price is to decrease the demand and \textit{not} to increase the revenue. To understand this point consider the $2^{nd}$ interval. The actual demand was $10$ VMs which could have lead to a revenue of $\$0.3$ if the rate was nominal. As the selling price increased to $\$0.045$ the demand reduced to $6$ VMs leading to an revenue of $\$0.27$. Therefore the net revenue in the $2^{nd}$ interval decreased even though the selling price increased. Similar argument is true for the $5^{th}$ interval. Therefore whenever there is an increase in price there is a decrease in revenue (and hence in profit) in that interval. Pictorially speaking, the revenue loss suffered in dynamic pricing can be captured by the area between the \textit{solid black} curve and the \textit{dashed blue} curve in Figure \ref{fig:motivating_example_graph}b.

\item[\textbf{2.}]Yet dynamic pricing makes more profit than static pricing. This is because in static pricing, many VMs are underutilized and hence does not contribute to the revenue. This is clearly shown in Figure \ref{fig:motivating_example_graph}c. Underutilized VMs for static case corresponds to the area between the \textit{solid black} curve and the \textit{dashed red} curve while that in dynamic case corresponds to the area between the \textit{dashed green} curve and the \textit{dashed blue} curve. Definitely the area corresponding to static case is more.

\item[\textbf{3.}]The idea behind dynamic pricing is: ``Suffer a small loss in one interval by decreasing the demand (Refered as \textit{Demand Loss} later) rather than buying a VM and then suffering a major loss in the subsequent intervals due to low demand (Refered as \textit{VM Loss} later)''.

\end{itemize}

\begin{figure}[t]
\begin{centering}
\includegraphics[scale=0.4]{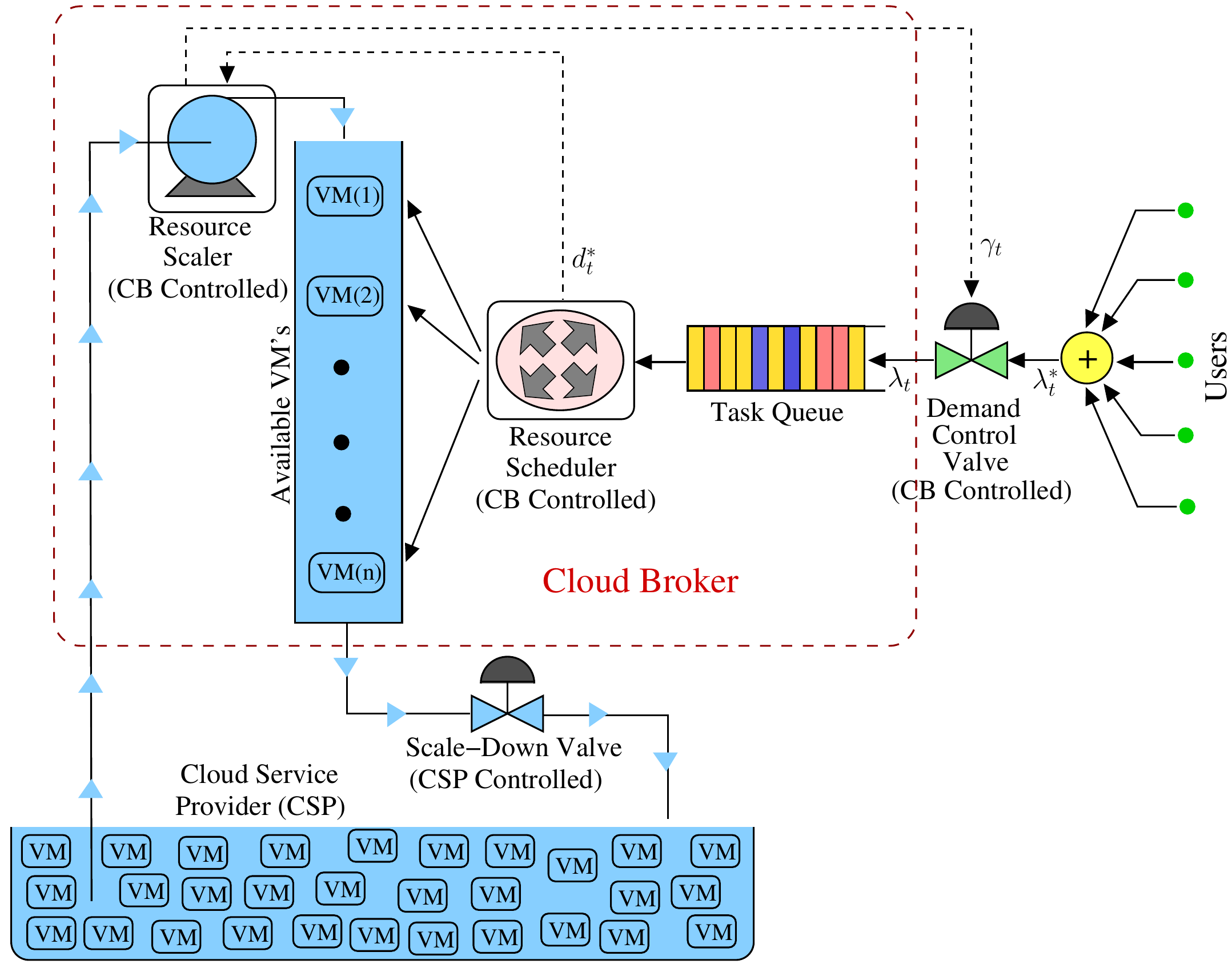}
\par\end{centering}

\caption{Cartoonistic representation of Cloud Brokerage mechanism.}

\label{fig:cloud_broker_arch}
\vspace{-0.7em}
\end{figure}

\subsection{Quantitative Modeling}
\label{sec:modeling}

Cloud brokerage mechanism basically consist of two blocks: 1) Resource Scheduler 2) Resource Scaler. This is shown in Figure \ref{fig:cloud_broker_arch}. Both resource scheduler and resource scaler are controlled by the cloud broker. The job of the Resource Scheduler is to schedule the incoming tasks onto the available VMs. While doing so it has to consider service level agreement (or SLAs). Resource Scheduler can be designed using well established theoretic tools as discussed in Section \ref{overview}. The role of the resource scaler is to rent (scale-up) VMs from CSP and also to perform dynamic pricing in order to maximize the profit of the cloud broker. In this paper, we assume that resource scheduler is already designed and concentrate on designing algorithms for Resource Scaler. Note that the design of Resource Scheduler and Resource Scaler can be decoupled. The Resource Scheduler should just update the Resource Scaler regarding the number of VMs required to complete the tasks in a given time slot. Before moving forward, please note the scale-down valve which is controlled by CSP. As shown in Figure \ref{fig:cloud_broker_arch}, this is not in control of the cloud broker. This aptly captures the notion of QBC and the process of \textit{smart-kill}. As mentioned in Section \ref{QBC}, due to the presence of QBC there is no point of giving back a VM before its billing cycle ends. This will be equivalent to the CSP automatically scaling down the VMs after its billing cycle. Therefore from cloud broker's perspective: ``Only scaling-up of VMs is controllable but scaling-down is not''.

We will now pose our problem mathematically. We consider that the user pays the cloud broker based on per-request/pay-what-you-use basis. In such a scenario the resource scaler has to solve the following profit optimization problem:

\vspace{-1.2em}

\begin{eqnarray*}
\mathbf{OP1} & : & \underset{\left\{ \gamma_{t},v_{t}\right\} }{\max}P=\overset{T}{\underset{t=1}{\sum}}\left(\gamma_{t}d_{t}-v_{t}\right)\\
\text{{subject}}\:\text{{to}} & : & \overset{t}{\underset{i=t-\tau+1}{\sum}}v_{i}\geq d_{t}\;;\forall t=1,2,\ldots,T\\
 &  & d_{t}=f\left(d_{t}^{*},\,\gamma_{t}\right)\;;\forall t=1,2,\ldots,T
\end{eqnarray*}

In optimization problem $\boldsymbol{OP1}$, $P$ is the profit to
be maximized. The term $\left(\gamma_{t}d_{t}-v_{t}\right)$ is the
profit at $t^{th}$ interval where, $\gamma_{t}$ is the selling price
per VM per time slot, $d_{t}$ is the number of VMs required to service the incoming
job request and $v_{t}$ is the number of VMs bought at $t^{th}$ interval. Without any loss of generality we normalize the cost price of a VM to $1$ unit. $\tau$ is the period of the billing cycle and hence $\overset{t}{\underset{i=t-\tau+1}{\sum}}v_{i}$
is the number of active VMs in the $t^{th}$ interval. $d_{t}$ is the modified VM demand when the
selling price is $\gamma_{t}$. The relation between the actual demand
$d_{t}^{*}$ and the modified demand $d_{t}$ is captured by the \textit{price-demand function}
$f\left(\cdot\right)$. If $\gamma_{t}=\gamma^{*}$, the nominal price, then $d_{t}=d^{*}_{t}$. The revenue earned by selling a VM at the nominal price of $\gamma^{*}$ for one complete billing cycle is $\gamma^{*} \tau$. For the cloud broker to make profit, $\gamma^{*} \tau>1$, $1$ being the cost price of a VM. It should be noted that all the variables associated with $\mathbf{OP1}$ lies in the set $\mathbb{R}_{+}$.

In conventional sense, the optimization problems dealt in ski-rental framework are minimization problems. Therefore we are interested in formulating $\mathbf{OP1}$ as an equivalent minimization problem. In this regard note that

\vspace{-1.2em}

\begin{eqnarray}
\label{eq9.1}
P & = & \overset{T}{\underset{t=1}{\sum}}\left(\gamma_{t}d_{t}-v_{t}\right)\nonumber \\
 & = & \overset{T}{\underset{t=1}{\sum}}\left[\gamma^{*}d_{t}^{*}-\left\{ \left(\gamma^{*}d_{t}^{*}-\gamma_{t}d_{t}\right)+v_{t}\right\} \right]\nonumber \\
 & = & \overset{T}{\underset{t=1}{\sum}}\left[\gamma^{*}d_{t}^{*}-\left\{ \left(\gamma^{*}d_{t}^{*}-\gamma_{t}\, f\left(d_{t}^{*},\,\gamma_{t}\right)\right)+v_{t}\right\} \right]
\end{eqnarray}

In equation \eqref{eq9.1} the first term $\gamma^{*}d_{t}^{*}$ is not controllable
and hence the maximization of $P$ becomes equivalent to minimization
of $\overset{T}{\underset{t=1}{\sum}}\left[\left(\gamma^{*}d_{t}^{*}-\gamma_{t}\, f\left(d_{t}^{*},\,\gamma_{t}\right)\right)+v_{t}\right]$.
We thus pose the following optimization problem which is equivalent
to $\boldsymbol{OP1}$:

\begin{eqnarray*}
\mathbf{OP2} & : & \underset{\left\{ \gamma_{t},v_{t}\right\} }{\min}L=\overset{T}{\underset{t=1}{\sum}}\left[\overset{Demand\: Loss}{\overbrace{\left(\gamma^{*}d_{t}^{*}-\gamma_{t}d_{t}\right)}}+\underset{VM\: Loss}{\underbrace{v_{t}}}\right]\\
\text{{subject}}\:\text{{to}} & : & \overset{t}{\underset{i=t-\tau+1}{\sum}}v_{i}\geq d_{t}\;;\forall t=1,2,\ldots,T\\
 &  & d_{t}=f\left(d_{t}^{*},\,\gamma_{t}\right)\;;\forall t=1,2,\ldots,T
\end{eqnarray*}

Intuitively speaking, $\mathbf{OP2}$ does the following: Consider that there is a hike in demand $d_{t}^{*}$ which decays soon. In such a case $\mathbf{OP2}$ will increase the selling price $\gamma_{t}$ to reduce the demand. In this way the cloud broker will suffer a small ``Demand Loss''. Buying enough VMs to support the demand hike is not a good option in such scenario as the cloud broker may suffer a huge ``VM Loss'' in subsequent intervals due to underutilized VMs. But if the hike in demand persists for a long time it is better to buy VMs to support this hike. However $\mathbf{OP2}$ is an offline optimization problem, i.e. to solve $\mathbf{OP2}$ we need $d_{t}^{*}$ for all $t=1,2,\ldots,T$. It is not possible to know in advance if an increase in demand is going to persist or will decay soon. The challenge is to design algorithms which can make such decisions online based on present and past data. Such algorithms are called online algorithms.

We now define the concept of Competitive Ratio which will be used later in Section \ref{ch:chap10} to compare the performance of the online algorithm with its optimal offline counterpart. Say that an online algorithm $A$ and the optimal algorithm $OPT$ ($\mathbf{OP2}$ here) suffers a loss $L_{A}\left(d^{*}\right)$ and $L_{OPT}\left(d^{*}\right)$ respectively for a given demand sequence $d^{*}=\begin{bmatrix}d_{1}^{*} & d_{2}^{*} & \cdots & d_{T}^{*}\end{bmatrix}^{T}$. Then algorithm $A$ is called $c-$competitive if:

\begin{equation}
\label{eq9.2}
L_{A}\left(d^{*}\right)\leq c\cdot L_{OPT}\left(d^{*}\right)\qquad\forall\, d^{*}\in\mathbb{R}_{+}^{T}
\end{equation}

Indeed $c\geq1$. In inequality \eqref{eq9.2} we have slightly misused the notation. $\mathbb{R}_{+}^{T}$ is a $T$ dimensional vector of non negative real numbers \textit{not} transpose of a non negative real number.

\subsection{Properties and Assumptions}
\label{assump}
\
$\;$

\noindent \textbf{Properties of Demand and Revenue Function}

$\;$

Most of the price demand function $f\left(d^{*},\,\gamma\right)$ (also called demand function) found in real world must satisfy the following conditions:

\begin{itemize}

\item[\textbf{1.}]$f\left(d^{*},\,\gamma\right) \geq 0$.

\item[\textbf{2.}]$f\left(d^{*},\,\gamma\right) \geq 0$ is monotonically increasing in $d^{*}$.

\item[\textbf{3.}]$f\left(d^{*},\,\gamma\right)$ is monotonically decreasing in $\gamma$ in the range $\left[\gamma^{*},\,\infty\right)$ while it is constant\footnote{We assume that the demand cannot increase above the actual demand $d^{*}$ even if the price decreases below $\gamma^{*}$} at $d^{*}$ in the range $\left[0,\,\gamma^{*}\right]$.

\item[\textbf{4.}]The revenue function $\gamma f\left(d^{*},\,\gamma\right)$, is monotonically decreasing in $\gamma$ in the range $\left[\gamma^{*},\,\infty\right)$. This captures the idea that increasing the selling price is to decrease the demand and \textit{not} to increase the revenue. As $f\left(d^{*},\,\gamma\right)$ is constant in $\left[0,\,\gamma^{*}\right]$, it is obvious that $\gamma f\left(d^{*},\,\gamma\right)$ will linearly increase in this range. Revenue function is maximum at $\gamma=\gamma^{*}$ and hence this is set as the nominal price.

\end{itemize}

Let $d=f\left(d^{*},\,\gamma\right)$. Then according to \textit{property 4} we have

\begin{equation}
\label{eq:new1}
-\infty<\frac{\partial}{\partial\gamma}\left(\gamma d\right)\leq0
\end{equation}

Differentiating inequality \eqref{eq:new1} using chain rule we get

\begin{equation}
\label{eq:new2}
-\infty<d+\gamma\frac{\partial d}{\partial\gamma}\leq0\;\Leftrightarrow\;-\infty<\frac{\partial d}{\partial\gamma}\leq-\frac{d}{\gamma}
\end{equation}

A demand function satisfying inequality \eqref{eq:new2} will be \textit{strictly} monotonically decreasing if $d>0$. Two cases may arise:

\textit{Case 1 (Semi-Infinite Operating Zone): }$d>0$ for all $\gamma \in \left[\gamma^{*} \; , \; \infty\right)$ and hence the demand function is strictly monotonically decreasing in this range.

\textit{Case 2 (Finite Operating Zone): }$d>0$ for all $\gamma \in \left[\gamma^{*} \; , \; \gamma_{op}\right)$. For $\gamma = \gamma_{op}$, $d=0$. Due to \textit{property 1}, $d=0$ for all $\gamma \geq \gamma_{op}$. Therefore the demand function is strictly monotonically decreasing in the range $\left[\gamma^{*} \; , \; \gamma_{op}\right)$.

\begin{figure}[t]
\begin{centering}
\includegraphics[scale=0.55]{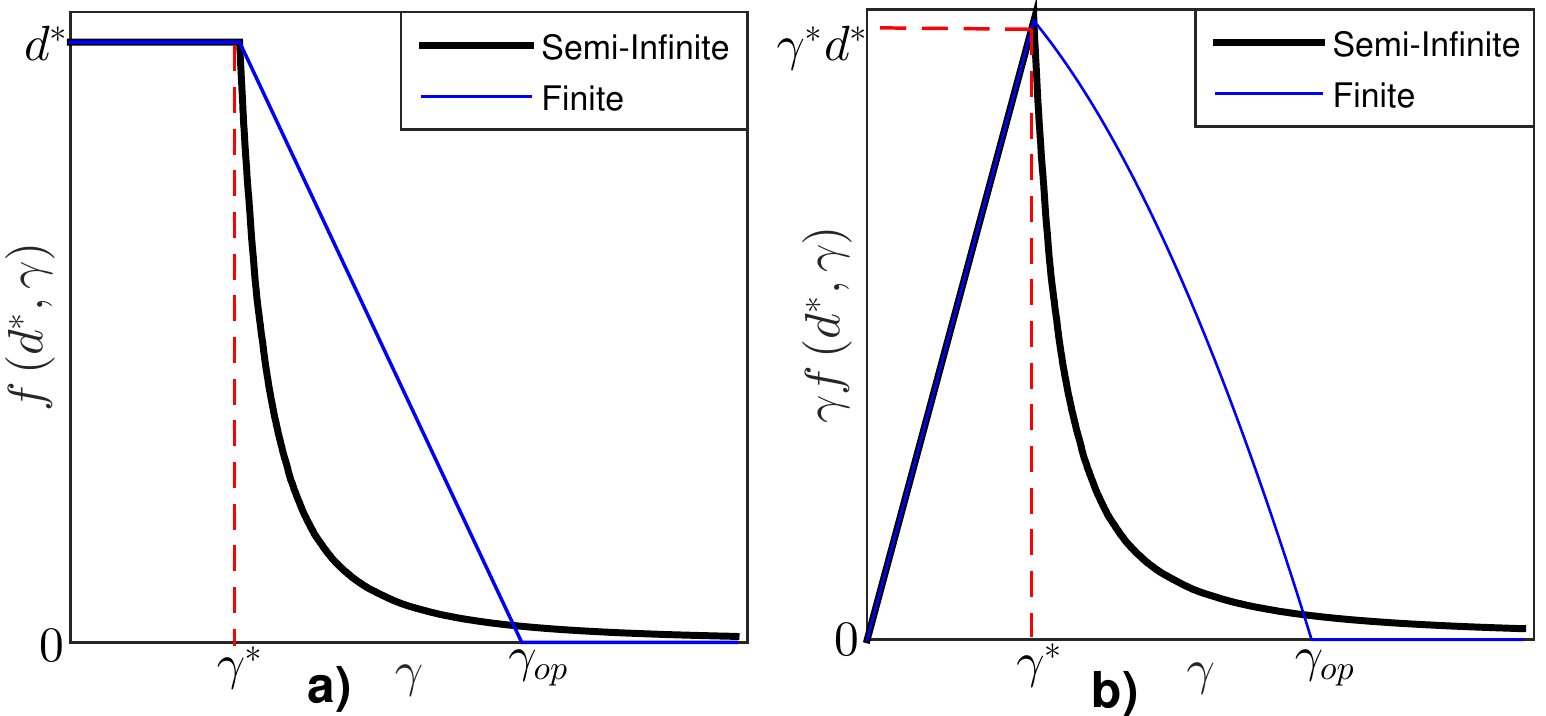}
\par\end{centering}

\caption{Graphs showing typical: a) Demand Functions b) Revenue Functions. Two demand functions and the corresponding revenue functions are shown. For the Semi-Infinite case $\left[\gamma^{*} \; , \; \infty\right)$ is the operating zone while for Finite case $\left[\gamma^{*} \; , \; \gamma_{op}\right)$ is the operating zone.}

\label{fig:PriceDemandFunction}
\end{figure}

The range of $\gamma$ where $f\left(d^{*},\,\gamma\right)$ is \textit{strictly} monotonically decreasing is called the \textit{operating zone}. Figure \ref{fig:PriceDemandFunction} illustrates the concept of operating zone for the above two cases. Strict monotonic nature of $f\left(d^{*},\,\gamma\right)$ in the operating zone implies that the function is invertible in this range. Mathematically, the following function exist in the operating zone,

\begin{equation}
\label{eq9.4}
\gamma=g\left(d^{*},\, d\right)
\end{equation}

The function $g\left( \cdot \right)$ returns the imposed price $\gamma_{t}$ given the actual demand $d_{t}^{*}$ and the modified demand $d_{t}$.

\textit{Remark: }The role of dynamic pricing is to regulate the demand. In the range $\left[0,\,\gamma^{*}\right]$, price has no effect on the demand and yet the cloud broker will incur a demand loss. Therefore to minimize $\mathbf{OP2}$ we work in the operating zone.

$\;$

\noindent \textbf{Assumptions in Problem Formulation}

$\;$

\begin{itemize}

\item[\textbf{1.}]We ignore the effect of \textit{reputation} while formulating our optimization problem. By increasing the price we force some tasks to exit the queue. By doing this we earn negative reputation of the users. A \textit{static} demand function of the form $d_{t}=f\left(d_{t}^{*},\,\gamma_{t}\right)$ does not capture the effect of pricing history and hence the role of reputation.

\item[\textbf{2.}]The partial online algorithm, discussed later in Section \ref{ch:chap10}, relies on demand prediction for future window $w$. In reality, only an estimate of future demand is possible however we assume perfect knowledge of future demand.

\item[\textbf{3.}]In real-life scenario, the relation between the modified demand $d_{t}$ and the actual demand $d_{t}^{*}$ is not governed by a deterministic function $f\left(\cdot\right)$. Rather the demand $d_{t}$ has a probability distribution\footnote{One may consider that the function $f\left(\cdot\right)$ is the mean of this probability distribution.} in the range $\left[0\,,\, d_{t}^{*}\right]$ for a given $\gamma_{t}$. However many works dealing with social welfare maximization using dynamic pricing (like \cite{ozdaglar,main3}) consider such deterministic demand function. We also assume the knowledge of $f\left(\cdot\right)$.

\item[\textbf{4.}]Define the following function in the operating zone,

\begin{equation}
\label{eq9.5}
p\left(d^{*},\, d\right)=\frac{\partial}{\partial d}\left[d\cdot g\left(d^{*},\, d\right)\right]
\end{equation}
\noindent Also define two more variables,

\begin{equation}
\label{eq9.6}
\begin{array}{rcl}
p_{m} & = & \underset{d^{*}\in\mathbb{R}_{+},\,0\leq d\leq d^{*}}{\min}\, p\left(d^{*}\,,\, d\right)\\
p_{M} & = & \underset{d^{*}\in\mathbb{R}_{+},\,0\leq d\leq d^{*}}{\max}\, p\left(d^{*}\,,\, d\right)
\end{array}
\end{equation}
\noindent We impose the following constrains on $p_{m}$ and $p_{M}$,
\begin{equation}
\label{eq9.7}
\frac{1}{\tau} < p_{m}\leq p_{M} < 1
\end{equation}

Inequality $p_{M} < 1$ implies that renting is cheaper than buying in short run while the inequality $\frac{1}{\tau} < p_{m}$ implies that buying is cheaper in the long run. We will further elaborate this in Section \ref{SkiRental}.

\end{itemize}

Let $\gamma = g\left(d^{*},\, d\right)$. Then according to inequality \eqref{eq9.7} we have

\begin{equation}
\label{eq:new3}
p_{m} \leq \frac{\partial}{\partial d}\left(\gamma d\right) \leq p_{M}
\end{equation}

Differentiating inequality \eqref{eq:new3} using chain rule we get

\begin{equation}
\label{eq:new4}
p_{m} \leq \gamma+d\frac{\partial\gamma}{\partial d} \leq p_{M}\;\Leftrightarrow\;\frac{p_{m}-\gamma}{d} \leq \frac{\partial\gamma}{\partial d} \leq \frac{p_{M}-\gamma}{d}
\end{equation}

According to inequality \eqref{eq:new2}, the following inequality is true in the operating zone

\begin{equation}
\label{eq:new5}
-\frac{\gamma}{d}\leq\frac{\partial\gamma}{\partial d}<0
\end{equation}

Inequality \eqref{eq:new4} and \eqref{eq:new5} is simultaneously satisfied if

\begin{equation}
\label{eq:new6}
\frac{p_{m}-\gamma}{d} \leq \frac{\partial\gamma}{\partial d}<\min\left(0\,,\,\frac{p_{M}-\gamma}{d}\right)
\end{equation}

This is because $\frac{p_{m}-\gamma}{d}>-\frac{\gamma}{d}$. However $\frac{p_{M}-\gamma}{d}\geq0$ for $\gamma\leq p_{M}$ while $\frac{p_{M}-\gamma}{d}<0$ for $\gamma>p_{M}$. It is trivial to observe that there exist a $\frac{\partial \gamma}{\partial d}$ satisfying inequality \ref{eq:new6} \textit{if and only if} $\gamma >p_{m}$. Also $\frac{\partial d}{\partial \gamma}$ is the slope of the demand function $f\left( \cdot \right)$ with respect to $\gamma$. Hence inequality \ref{eq:new6} upper and lower bounds the slope of the demand function for a given $d$ and $\gamma$.

\section{Online Optimization Problem}
\label{ch:chap10}

\subsection{Ski-Rental Problem}
\label{SkiRental}

The ski-rental problem abstracts a class of problem in which a player has to decide whether to buy or rent a resource without a priori knowledge of the period of usage. Renting is cheaper if the period of usage is short while buying is cheaper in the long run. In the original problem a skier is faced with the option of either buying or renting a set of skis without knowing in advance the number of days she will be skiing.

Cost of buying skis is $\$ 1$ while renting cost $\$ P$ per day where $P<1$. If the skier knows in advance that she will be skiing for $y$ days then the choice of buying or renting is simple. If $y \geq \frac{1}{P}$ then the skier will buy the skis in the very first day. Otherwise she will keep renting the skis for $y$ days.

The online case is more challenging. In ski-rental literature, the concept of \textit{breakeven point} is used to design online algorithms. Such algorithms suggest that the skier should keep on renting the skis till the $n^{th}$ day when the cost of renting $nP$, is more than the cost of buying, i.e. $nP>1$. On the $n^{th}$ day she should buy the skis. These is shown to be the most optimal \textit{deterministic} online algorithm and has a competitive ratio of $2$.

Ski-Rental problem has been used to solve real life problems like TCP acknowledgement problem \cite{algorithmica}, Bahncard problem \cite{bahncard} etc. Its application in cloud computing has already been discussed in Section \ref{sec:contribution}. The key step towards using ski-rental literature for our problem would be to map the following four entities in our context: \textbf{1)} Renting \textbf{2)} Buying \textbf{3)} Buying Cost \textbf{4)} Renting Cost. In remaining of this section we will define these entities.

\textbf{1. Renting: }It is the process of decreasing the demand by increasing the selling price of the VMs. To decrease actual demand $d_{t}^{*}$ to modified demand $d_{t}$ we impose selling price of $\gamma_{t}=g\left(d_{t}^{*},\, d_{t}\right)$.

\textbf{2. Buying: }It is the process of buying $v_{t}$ VMs to support the modified demand $d_{t}$. If the number of active VMs in the beginning of time slot $t$ is $x_{t}$ then $v_{t}=\max\left(0,\, d_{t}-x_{t}\right)$.

\begin{figure}[t]
\begin{centering}
\includegraphics[scale=0.55]{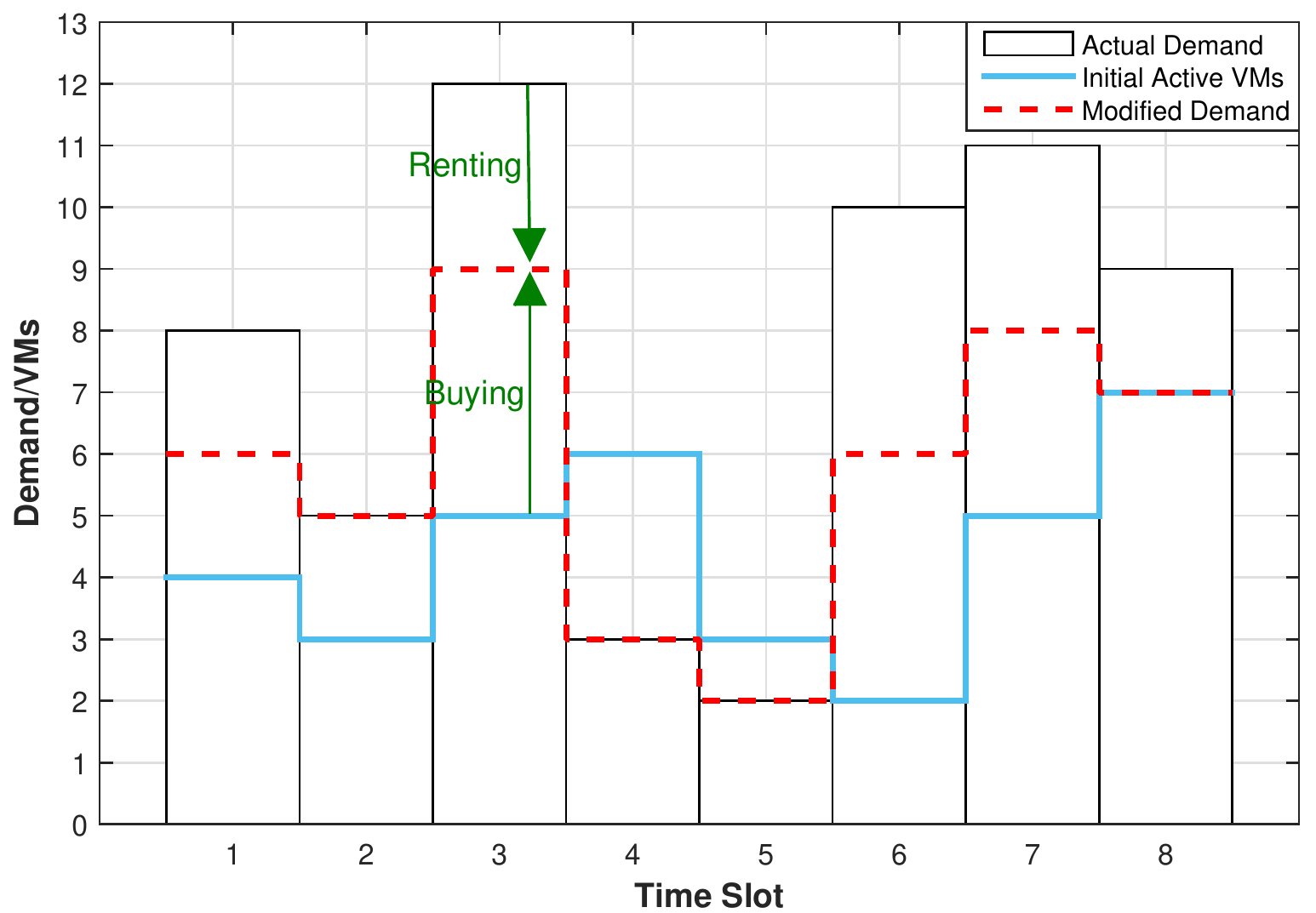}
\par\end{centering}

\caption{Figure showing the actual demand $d_{t}^{*}$ (solid black curve), modified demand $d_{t}$ (dashed red curve) and the number of active VMs $x_{t}$ in the beginning of every time slot (solid blue curve).}

\label{fig:DynamicPricing_SkiRental}
\end{figure}

Figure \ref{fig:DynamicPricing_SkiRental} illustrates the renting and buying process. In the $3^{rd}$ time slot renting is equivalent to reducing the demand from $12$ to $9$ while buying is the process of purchasing $4$ VMs. Similarly in the $1^{st}$, $6^{th}$ and the $7^{th}$ time slot there is both renting and buying. There is no renting in the $2^{nd}$ time slot, we only buy $2$ VMs. On contrary there is no buying in the $8th$ time slot, we only rent the demand from $9$ to $7$. In the $4^{th}$ and the $5^{th}$ time slot there is neither renting nor buying because the number of active VMs in the beginning of these time slots is more than the actual demand.

\textbf{3. Buying Cost: }It is the cost of buying $n$ VMs. As the cost price of VM is assumed to be $1$ unit, the cost of buying $n$ VMs is equal to $n$ units.

\textbf{4. Renting Cost: }It is the demand loss in a given time slot associated with reducing a demand from $d$ to $d-n$, where $n \leq d$, when the actual demand is $d^{*}$. Mathematically
\vspace{-0.3em}
\begin{equation}
\label{eq10.1}
R\left(d^{*},\,d,\,n\right)=g\left(d^{*},\, d\right) \cdot d-g\left(d^{*},\, d-n\right) \cdot \left(d-n\right)
\end{equation}

It should be noted that unlike the renting cost found in other ski-rental literature, for our case, $R$ is not a constant. It is a function of $d^{*}$, $d$ and $n$. However we will not mention these parameters explicitly for notational simplicity. \textit{Please Note: }``Renting Cost" and ``demand loss" means the same and will be used interchangeably. 

Now we will explain the importance of inequality constrain \eqref{eq9.7}. As mentioned before renting cost should be more than the buying cost in the long run. Given that the billing cycle is of $\tau$ period, the cost of renting for $\tau$ period should be greater than the cost of buying. Otherwise buying of VMs will never be required. The cost of buying $n$ VMs is $n$ while the cost of Renting of $n$ demands for $\tau$ period is $R\tau$. Hence, $R\tau \geq n$. Similarly renting cost should be lesser than the buying cost in the short run. Therefore the cost of renting $n$ demands for $1$ period should be less than the cost of buying $n$ VMs. Hence, $R \leq n$. Therefore to formulate our problem in Ski-Rental framework, it is necessary that the following inequality holds

%
%
%
\begin{equation}
\label{eq10.2}
\frac{n}{\tau} \leq R \leq n
\end{equation} 

\noindent \textbf{Proposition 1: }If inequality \eqref{eq9.7} is satisfied then inequality \ref{eq10.2} will hold true.

\noindent \textbf{Proof: }Consider the following,
\vspace{-0.5em}
\begin{eqnarray}
R & = & g\left(d^{*},\, d\right) \cdot d-g\left(d^{*},\, d-n\right) \cdot \left(d-n\right)\nonumber \\
 & = & \int_{d-n}^{d}\,\frac{\partial}{\partial \theta}\left[\theta \cdot g\left(d^{*},\, \theta \right)\right]\: d\theta\nonumber \\
 & = & \int_{d-n}^{d}\, p\left(d^{*},\, \theta\right)\: d\theta\\
 & \geq & \int_{d-n}^{d}\, p_{m}\: d\theta=p_{m}n
\end{eqnarray}

\noindent Equation (15) and inequality (16) comes from the definition of $p\left(d^{*},\, x\right)$ and $p_{m}$ (refer equation \eqref{eq9.5} and \eqref{eq9.6}). Similarly,
\begin{equation}
\label{eq10.5}
R \leq p_{M}n
\end{equation}
\noindent The qualitative interpretation of inequality (16) and inequality (17) is that the minimum and maximum renting cost of $1$ demand is $p_{m}$ and $p_{M}$ respectively. If inequality \eqref{eq9.7} holds, then $\frac{1}{\tau} \leq p_{m}$. Substituting this in inequality (16) we get
\begin{equation}
\label{eq10.6}
\frac{n}{\tau} \leq R
\end{equation}
\noindent Similarly if inequality \eqref{eq9.7} holds, then $p_{M} \leq 1$. Substituting this in inequality (17) we get
\begin{equation}
\label{eq10.7}
R \leq n
\end{equation}
\noindent Combining inequality \eqref{eq10.6} and \eqref{eq10.7} we get inequality \eqref{eq10.2}. This completes the proof.

\subsection{Online Algorithm}
\label{Online}

As discussed earlier in the previous subsection the concept of breakeven point has been widely used to design online algorithms in ski-rental literature. The key idea is to keep renting till the time renting cost equals buying cost. Here buying cost is the breakeven point. If renting cost exceeds the buying cost, we buy the resource. In this section we will apply these concepts to design two online algorithm: 1) Fully Online Algorithm which has no knowledge of future demand 2) Partial Online Algorithm, which assumes perfect future demand information for future window $w<\tau$. Intuitively speaking, both Fully Online Algorithm and Partial Online Algorithm works in \textit{pessimistic sense}. It always assumes that an increase in demand is not going to persist and hence it reduces the demand by increasing the selling price of the VMs. This reduction in demand incurs a renting cost. However at every time interval it calculates the Net Renting Cost in the past and the future intervals. If the Net Renting Cost exceeds $1$ unit (the buying cost of $1$ VM), it buys a new VM.

The psuedocode of partial online algorithm with future window of $w$ is given in Algorithm \ref{algo:partialonline}. The same psuedocode is applicable for the fully online algorithm if we substitute $w=0$. Both fully/partial online algorithm consist of three basic steps. In the following we will explain these steps in relation to the fully online algorithm.

\begin{algorithm}[t]
Let $x_{i}$ be the number of VM's at time $i$

\textbf{1.} Set $x_{i}=0\,;\, i=1,2,\ldots,T$ 

\textbf{2.} Predict actual demand $d_{i}^{*}$ for $i=t,\, t+1,\,\ldots,\, t+w$.

\textbf{3.} $do$

\textbf{4.} $\quad$Set Net Renting Cost $l=0$. Also set $i=t+w-\tau+1$.

\textbf{5.} $\quad$ $while\:\left(i\leq t+w\right)$

\textbf{6.} $\qquad$ $if\:\left(x_{i}+1\leq d_{i}^{*}\right)$

\textbf{7.} $\qquad\;\;\:$ $l=l+\left[g\left(d_{i}^{*},\, x_{i}+1\right)\cdot\left(x_{i}+1\right)-g\left(d_{i}^{*},\, x_{i}\right)\cdot x_{i}\right]$

\textbf{8.} $\qquad$ $end\: if$

\textbf{9.} $\qquad$ $i=i+1$

\textbf{10.} $\:\:$ $end\: while$

\textbf{11.} $\:\:$ $if\:\left(l\geq1\right)$

\textbf{12.} $\qquad$Buy a new VM: $v_{t}=v_{t}+1$.

\textbf{13.} $\qquad$Update the number of VM's that can be used in future:

$\qquad\quad\;\;$ $x_{i}=x_{i}+1\,;\, i=t,\, t+1,\ldots,t+\tau-1$

\textbf{14.} $\qquad$Update the number of VM's in the history indicating

$\qquad\quad\;\;$that previous mistakes have been corrected:

$\qquad\quad\;\;$ $x_{i}=x_{i}+1\,;\, i=t+w-\tau+1,\ldots,t-1$

\textbf{15.} $\:\:\;$ $end\: if$

\textbf{16.} $while\:\left(l\geq1\right)$

\textbf{17.} $if\:\left(x_{t}\leq d_{t}^{*}\right)$

\textbf{18.} $\quad$ $\gamma_{t}=g\left(d_{t}^{*},\, x_{t}\right)$

\textbf{19.} $else$

\textbf{20.} $\quad$ $\gamma=\gamma^{*}$

\textbf{21.} $end\: if$

\textbf{22.} Jump to Step 2.

\protect\caption{Partial Online Algorithm with future prediction window $w$}
\label{algo:partialonline}
\end{algorithm}

\noindent \textbf{Step 1. Calculating Net Renting Cost}

We calculate the net renting cost $l$ which we could have been saved if we rented $1$ less demand in each slot for the past $\tau$ period. Let $x_{i}$ be the number of active VMs at time $i$. Then the net renting cost is

\begin{equation}
\label{eq10.8}
l=\sum_{i=t-\tau+1}^{t}R\left(d^{*},\, x_{i}+1,\,1\right)
\end{equation}

Equation \eqref{eq10.8} is valid only if $\ensuremath{x_{i}+1\leq d_{i}^{*}}\,;\:\forall i\in\left[t-\tau+1,\, t\right]$. If for a given interval $x_{i}+1 > d_{i}^{*}$, the renting cost for that interval is $0$. This is done in Steps 4 to 10 of Algorithm \ref{algo:partialonline}.

\noindent \textbf{Step 2. Buying new VM}

This is done in Steps 11 to 16 of Algorithm \ref{algo:partialonline}. In our case the cost of a VM is $1$. If $l \geq 1$ then the corresponding demands should not have been reduced. Rather we should have bought a VM to serve them. To compensate for this mistake we buy a VM in the current time slot. We increase the current and the future $x_{i}$ by $1$ to indicate that an extra VM is available. We also increase the past $x_{i}$ by $1$ to indicate that a corrective measure was taken. We then jump back to Step 1. However if $l < 1$, we jump to Step 3.

\noindent \textbf{Step 3. Setting the Selling Price}

Let the number of active VMs in the current time slot be $x_{t}$ after performing Step 2 and Step 3. Then $x_{t}$ is the modified demand. So we set our selling price as $\gamma_{t}=g\left(d^{*}_{t},x_{t}\right)$.

$\quad$

The partial online algorithm is almost same as the fully online algorithm. The difference lies in the calculation of Net Renting Cost in \textbf{Step 1}. In case of fully online algorithm it is calculated for the period $t-\tau+1$ to $t$ while for partial online algorithm it is calculated for the period $t+w-\tau+1$ to $t+w$.

$\quad$

\noindent \textbf{Theorem 1: }Competitive Ratio of partial online algorithm is
\begin{equation}
\label{eq10.9}
c\left(\alpha\right)=1+\min\left(1\,,\, p_{M}\tau\left(1-\alpha\right)\right)
\end{equation}

\noindent where $\alpha=\frac{w}{\tau}$ and $w < \tau$.

\noindent \textbf{Proof: }Please refer appendix for the proof.

$\quad$

\noindent \textbf{Corollary 1: }Fully online algorithm is $2$-competitive.

\noindent \textbf{Proof: }For fully online algorithm $\alpha=0$.
Hence $c=1+\min\left(1\,,\, p_{M}\tau\right)$. According to inequality \eqref{eq9.7} we have $p_{m}\tau\geq1$. Also $p_{M}\geq p_{m}$ implying $p_{M}\tau\geq1$.
Therefore $\min\left(1\,,\, p_{M}\tau\right)=1$ and hence $c=2$.

\begin{figure}[t]
\begin{centering}
\includegraphics[scale=0.4]{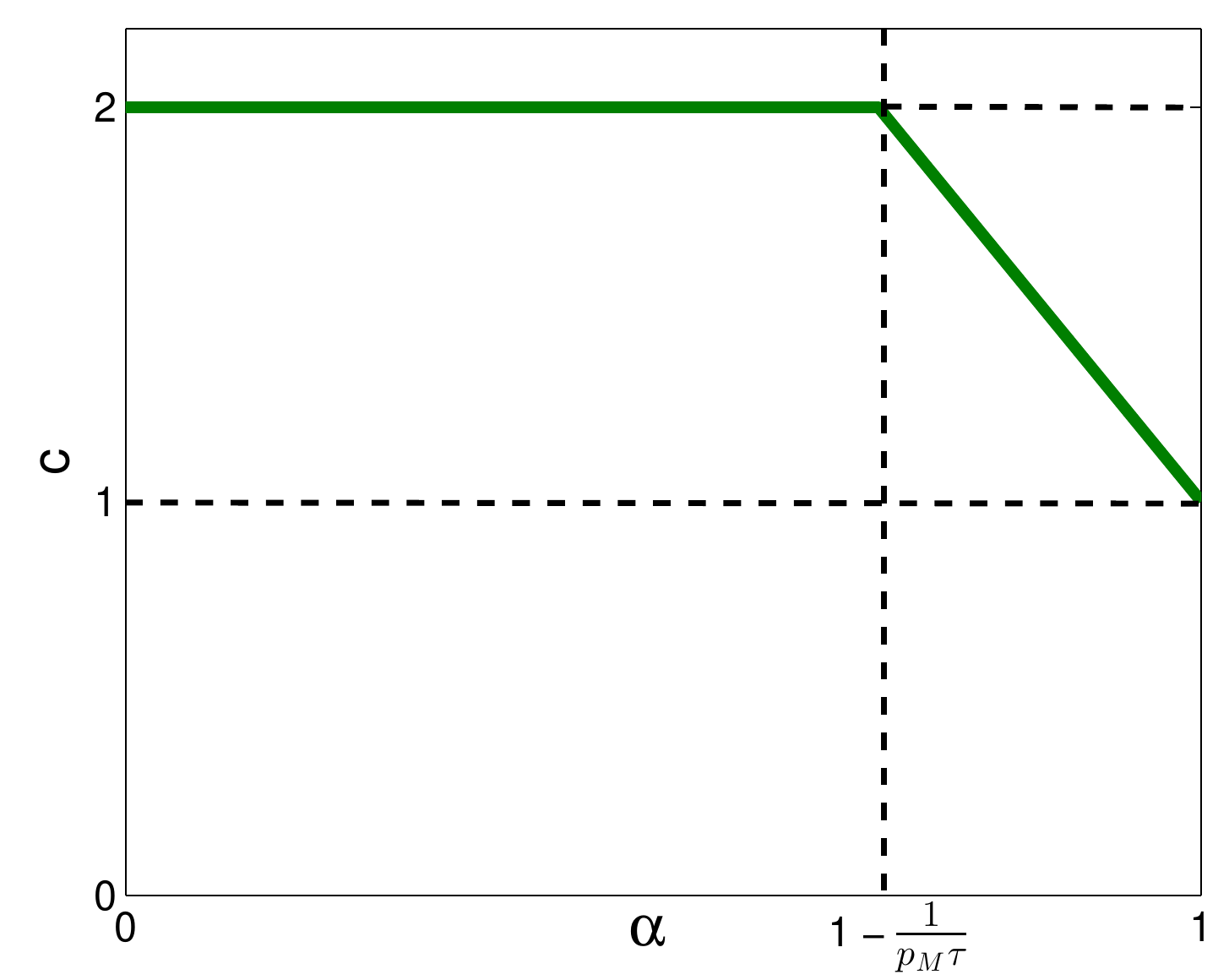}
\par\end{centering}

\centering{}\protect\caption{Graph showing the competititve ratio of partial online algorithm.}
\label{fig:partialcompratio}
\end{figure}

Note that the competive ratio of the partial online algorithm
can be more explicitly written as
\vspace{0.5em}
\[
c=\begin{cases}
2 & ;\;\alpha\leq1-\frac{1}{p_{M}\tau}\\
1+p_{M}\tau\left(1-\alpha\right) & ;\;\alpha>1-\frac{1}{p_{M}\tau}
\end{cases}
\]
\vspace{0.5em}
\noindent As $p_{M}\tau>1$, there always exist an $\alpha_{M}\in\left(0,\,1\right)$
such that $1+p_{M}\tau\left(1-\alpha\right)<2;\;\forall\alpha>\alpha_{M}$.
This gives the theoretical guarantee that future demand information
indeed improves the performance of the online algorithm. Figure \ref{fig:partialcompratio} shows a typical plot of equation \eqref{eq10.9}.

\section{Simulation Results}

We performed simulations driven by real world traces to validate the online dynamic pricing algorithm proposed in the paper.

The first step is to generate the actual demand curve. To do this we have used google cluster usage traces available in \cite{googlecluster}. The actual demand curve is shown in Figure 7b and 7d. The curve spans $1$ day and is slotted in $5$ min interval. Cost price of a VM is taken to be $1$ unit. A VM has a billing cycle of $1$ hour and hence $\tau=12$. The next step is to generate the demand function. A real world demand function can only be inferred by doing a market survey. But for the sake of simulation, we have synthesized our own demand function. This can be explained in steps:

\begin{itemize}

\item[\textbf{1.}]We consider a demand function of the form $\frac{d}{d^{*}}=f\left( \gamma \right)$.

\item[\textbf{2.}]A value of $p_{m}$ and $p_{M}$ satisfying inequality \eqref{eq9.7} is chosen.

\item[\textbf{3.}]A finite interval of price $\left[\gamma^{*} \, , \, \gamma_{o} \right]$ is uniformly divided into small parts.

\item[\textbf{4.}]For each part we substitute the corresponding value of $\gamma$ and $d$ in inequality \eqref{eq:new6}. The value of $\frac{\partial\gamma}{\partial d}$ for this part is chosen in random such that it satisfies inequality \eqref{eq:new6}. Using this value of $\frac{\partial\gamma}{\partial d}$, $d$ for the next part is calculated.

\end{itemize}

As part of this simulation we will conduct two comparative studies, first to study the effect of demand prediction and secondly the effect of $p_{m}$.

To study the effect of demand prediction we first synthesized a demand function with $p_{m}=\frac{1}{12}$ and $p_{M}=0.8$. We then simulated \textit{Algorithm 1} for $w=0$ and $w=4$. The results of the simulation is clearly shown in Figure 7b and 7c. Compared to $w=4$, the reduction in demand is more in $w=0$. This shows the pessimistic nature of our online algorithm, i.e. it prefers reducing the demand compared to buying new VMs. With increase in future window $w$ the algorithm tends towards the optimal counterpart. The net profit $P=\sum_{t=1}^{T}\left(\gamma_{t}d_{t}-v_{t}\right)$ for $w=0$ is $781$ units while for $w=4$ it is $937$ units. The net profit $P$ is more for $w=4$ and hence the net loss $L$ (refer $\mathbf{OP2}$) is less. This is in consensus with \textit{Theorem 1}.

We next studied the effect of $p_{m}$ on our algorithm. To do this we constructed another demand function with $p_{m}=\frac{3}{12}$ and $p_{M}=0.8$. From Figure 7d and 7e we can observe that if $p_{m}$ is high the effect of pricing on demand is less. This is an obvious consequence of the definition of $p_{m}$.

\begin{figure*}
\label{fig:simfinal}
\begin{centering}
\includegraphics[scale=0.6]{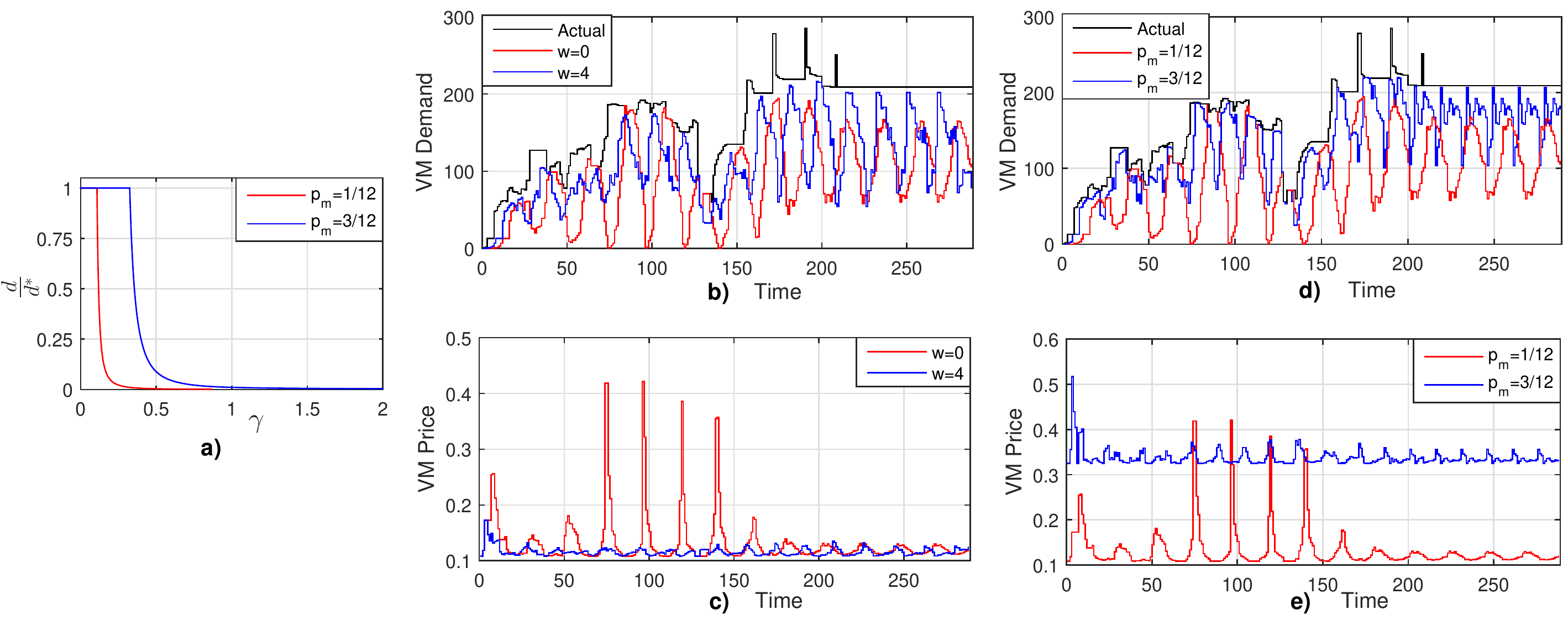}
\par\end{centering}

\protect\caption{a) Demand function for two different value of $p_{m}$. b), c) Plots of actual and modified demands and the corresponding pricing to compare the effect of prediction window $w$. d), e) Plots of actual and modified demands and the corresponding pricing to compare the effect of $p_{m}$.}
\end{figure*}

\section{Discussion and Extensions}
\label{ch:chap11}
We discussed the unique challenge posed by the presence of quantized billing cycles. The dynamic pricing strategy proposed in this paper can be considered as an alternative to the work done in \cite{main1} to maximize the profit of the cloud broker. Merging our algorithm with that of \cite{main1} should be very interesting. Such a merging will lead to a very interesting class of problems where the cloud broker has to decide whether to \textbf{a)} reduce the demand by increasing the VM price \textbf{b)} buy \textit{on-demand} VMs to support the demand \textbf{c)}  reserve VMs. This problem is similar to multislope ski-rental problem.

Two \textit{deterministic} online algorithms were designed to increase the profit of cloud brokers in the presence of QBC. The competitive ratio of both the algorithms were derived. We showed the importance of demand prediction by deriving a better competitive ratio for the partial online algorithm than those found in ski-rental literature. In similar lines we would like to explicitly point out that the competitive ratio of \textit{Algorithm 3} of \cite{main1}, i.e. deterministic algorithm with demand prediction window $w$, has a competitive ratio of $1+p \tau \min \left( 1\; , \; \frac{w}{\tau}\right)$ which is better than that reported in \cite{main1} (refer \textit{Proposition 5}). This result is new in ski-rental literature.

It has been widely reported in ski-rental literature that \textit{randomized} algorithms has better competitive ratio than its \textit{deterministic} counterpart. Extending our algorithm to its randomized counterpart should be trivial. However deriving a better competitive ratio for the randomized algorithm with demand prediction may be challenging.

The key idea of our algorithms is to use pricing signal to regulate user demand. One may argue that such an algorithm gives poor service to the user as it pushes tasks out of the queue in order to maximize cloud broker's profit. We would like to make few comments in this regard:
\begin{itemize}

\item [\textbf{1.}]Those tasks which gets pushed out of the queue can enter it again at a later instant. So we are not rejecting the tasks. Rather we are deferring it. Our algorithm is specifically good if $p_{M}$ is low. A low $p_{M}$ implies that even with a small increase in price there will be significant change in demand. Qualitatively, a low $p_{M}$ correspond to tasks with low priority. Such tasks can be deferred.

\item [\textbf{2.}]Some literature considers penalizing the cloud provider when it fails to meet SLA's (refer \cite{penalize}. Such a practice is not encouraged in cloud computing as it is a service oriented computing paradigm. If the cloud provider accepts the task, it must meet the SLAs. If it cannot it is better that it rejects the task. We are following similar practice.

\item [\textbf{3.}]We can penalize cloud broker to push tasks out of the queue by introducing \textit{reputation factor}. As discussed in section \ref{assump} we need to consider dynamic price-demand function to include reputation factor. Such an extension of our work will be challenging but a very fruitful research.

\end{itemize}

As an immediate extension of our work we are interested in two areas. \textit{Assumption 4} discussed in Section \ref{assump} may not be true for all real world demand functions. We are exploring the implication of relaxing it. Our intuition is that if we make a small modification in \textit{Algorithm 1}, the competitive ratio will remain the same. We are also actively working on the randomized algorithm for this problem. More specifically we are exploring the possibility of improving the competitive ratio of the randomized algorithm in the presence of statistical information about the actual demand. This is motivated by the work done in \cite{constrained_skirental}.

\section*{APPENDIX}
\label{comp_ratio}

We will denote the partial online algorithm with future window of $w$ by $A_{w}$ and the offline optimal algorithm $\mathbf{OP2}$ by $OPT$. We will first consider two lemmas which are important for the proof.

\noindent \textbf{Lemma 1: }Let $A_{w}$ buy $N_{w}$ VMs while $OPT$ buy $N_{OPT}$ VMs throughout the time duration of $t=1$ to $t=T$. Then for any demand sequence $d_{t}^{*}$, $N_{OPT} \geq N_{w}$.

\textit{Lemma 1} is an obvious consequence of the pessimistic nature of the fully/partial online algorithm as discussed in Section \ref{Online}. Both these algorithms assumes that a rise in demand is not going to persist and hence has the tendency to reduce the demand by increasing the selling price rather than buying VMs to support the rise in demand. Therefore $N_{OPT} \geq N_{w}$. Please refer Appendix A of \cite{main1} for the proof.

\noindent \textbf{Lemma 2: }The net renting cost of those demands that were served by the same VM in $OPT$ should be greater than or equal to $1$ (the breakeven point).

\noindent \textbf{Proof: }The proof directly follows from the definition of $OPT$. We prove this lemma by contradiction. Consider that in $OPT$ a VM was bought to serve demands whose net renting cost is less than $1$. The loss suffered to buy a VM to support these demands is more than the loss suffered to rent these demands. Therefore there can be a better algorithm to reduce the loss which contradicts the definition of $OPT$.

Let $A_{w}$ and $OPT$ rent $n_{t}$ and $N_{t}$ demands respectively at time $t$. Let $\mathcal{R}\left(A_{w}\right)$ and $\mathcal{R}\left(OPT\right)$ be the net renting cost of $A_{w}$ and $OPT$ respectively. We have,

\vspace{-1.0em}

\[
\mathcal{R}\left(A_{w}\right)=\sum_{t=1}^{T}R\left(d_{t}^{*},\, d_{t}^{*},\, n_{t}\right)
\]

\[
\mathcal{R}\left(OPT\right)=\sum_{t=1}^{T}R\left(d_{t}^{*},\, d_{t}^{*},\, N_{t}\right)
\]

Let $\mathcal{R}\left(A_{w} \backslash OPT\right)$ denote the net renting cost incurred in $A_{w}$ which is not incurred in $OPT$. Mathematically

\vspace{-1.0em}

\[
\mathcal{R}\left(A_{w}\backslash OPT\right)=\sum_{t=1}^{T}\left[R\left(d_{t}^{*},\, d_{t}^{*},\, n_{t}\right)-R\left(d_{t}^{*},\, d_{t}^{*},\, N_{t}\right)\right]^{+}
\]

\noindent where $\left( x \right )^{+} = \max \left( 0,x\right)$. We are interested in upper bounding $\mathcal{R}\left(A_{w} \backslash OPT\right)$. To do this, consider the demands that were rented in $A_{w}$ but not in $OPT$. These demands were served by not more than $N_{OPT}$ VMs in $OPT$. We focus on the demands served by one of these $N_{OPT}$ VMs. This is shown in the following figure

\begin{figure}[t]
\begin{centering}
\includegraphics[scale=0.35]{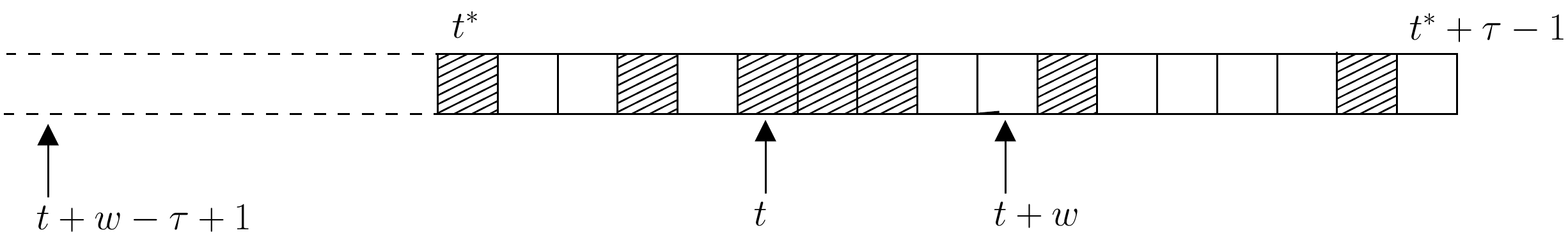}
\par\end{centering}

\caption{Figure showing the demands served by a VM in $OPT$.}

\label{fig:VMstrip}
\vspace{-1.0em}
\end{figure}

Figure \ref{fig:VMstrip} shows the demands served by a VM in $OPT$ which is bought at time $t^{*}$ and last till $t^{*}+ \tau -1$. The shaded areas shows the demands which are served by this VM. If the area is not shaded then in that time slot the VM does not serve any demand. The shaded areas can be scattered anywhere in the time interval $\left[ t^{*},\; t^{*}+ \tau -1 \right]$. According to \textit{Lemma 1}, the net renting cost of the demands depicted by these shaded areas should be greater than or equal to $1$.

Now we investigate how these demands will be treated by $A_{w}$. While calculating the net renting cost, $A_{w}$ considers both future demand and past demand. This is shown in figure \ref{fig:VMstrip} by the three arrows. $A_{w}$ calculates the net renting cost in the interval $\left[ t+w - \tau +1,\; t+w \right]$ where $t$ is the current time, the interval $\left[ t+w - \tau +1,\; t \right]$ corresponds to the past window while $\left[ t+1,\; t+w \right]$ is the future window. If the net renting cost in the interval $\left[ t+w - \tau +1,\; t+w \right]$ becomes $1$, then $A_{w}$ will buy a VM. As mentioned before the net renting cost in the interval $\left[ t^{*},\; t^{*}+ \tau -1 \right]$ is definitely greater than or equal to $1$ (due to \textit{Lemma 1}). Therefore there is definitely a time $t$ satisfying $t^{*} \leq t \leq t^{*} + \tau -w$ when $A_{w}$ will buy a VM. The demands in the interval $\left[t^{*},\; t\right]$ will be rented while that in the interval $\left[t+1,\; t^{*} + \tau -1\right]$ will be served by the VM. The maximum renting cost of the demands in the interval $\left[t^{*},\; t\right]$ is $\min\left(1\,,\, p_{M}\left(t-t^{*}\right)\right)$. This can be understood as follows:

\begin{itemize}

\item[\textbf{1.}]The maximum renting cost of one demand is $p_{M}$ (refer inequality (17)). If all the intervals in the past window $\left[t^{*},\; t\right]$ has a demand, then the maximum renting cost in the past window is $p_{M}\left(t-t^{*}\right)$. However the renting cost in the interval $\left[t^{*},\; t+w \right]$ is equal to $1$ and hence the renting cost in the interval $\left[t^{*},\; t\right]$ cannot exceed $1$. Therefore the maximum renting cost in the interval $\left[t^{*},\; t\right]$ is $\min\left(1\,,\, p_{M}\left(t-t^{*}\right)\right)$.

\item[\textbf{2.}]Due to inequality \eqref{eq9.7}, there \textit{may be} a $t$ satisfying $t^{*} \leq t \leq t^{*} + \tau -w$ such that $p_{M}\left(t-t^{*}\right)>1$.

\end{itemize}

Therefore if $A_{w}$ buys a VM at time $t$ then the demands in the interval $\left[t^{*},\; t\right]$ was rented in $A_{w}$ while it is served by a VM in $OPT$. The maximum renting cost of these demands is $\min\left(1\,,\, p_{M}\left(t-t^{*}\right)\right)$. As mentioned before, $t \leq t^{*} + \tau -w$. Therefore the maximum renting cost of those demands which are served by a VM in $OPT$ but rented in $A_{w}$ is upper bounded by $\min\left(1\,,\, p_{M}\left( \tau - w \right)\right)$. Given that at most $N_{OPT}$ VMs serve such demands we have

\vspace{-1.0em}

\begin{eqnarray}
\label{eqD.1}
\mathcal{R}\left(A_{w}\backslash OPT\right) & \leq & \min\left(1\,,\, p_{M}\left(\tau-w\right)\right)N_{OPT}\nonumber \\
 & = & \min\left(1\,,\, p_{M}\tau\left(1-\alpha\right)\right)N_{OPT}
\end{eqnarray}

\noindent where $\alpha=\frac{w}{\tau}$. Let $\delta_{M}=\min\left(1\,,\, p_{M}\left(\tau-w\right)\right)$. We will use inequality \eqref{eqD.1} to upper bound $\mathcal{R}\left(A_{w}\right)$:

\vspace{-1.0em}

\begin{eqnarray}
\label{eqD.2}
\mathcal{R}\left(A_{w}\right) & \leq & \mathcal{R}\left(OPT\right)+\mathcal{R}\left(A_{w}\backslash OPT\right)\nonumber \\
 & \leq & \mathcal{R}\left(OPT\right)+\delta_{M} N_{OPT}
\end{eqnarray}

Let $L_{OPT}$ be the net loss incurred in $OPT$. Then,

\vspace{-1.0em}

\begin{equation}
\label{eqD.3}
L_{OPT}=\mathcal{R}\left(OPT\right)+N_{OPT}\geq N_{OPT}
\end{equation}

Similarly let $L_{A_{w}}$ be the net loss incurred in $A_{w}$. We have,

\vspace{-1.0em}

\begin{eqnarray}
L_{A_{w}} & = & \mathcal{R}\left(A_{w}\right)+N_{w}\\
 & \leq & \mathcal{R}\left(OPT\right)+\delta_{M} N_{OPT}+N_{w}\\
 & \leq & \min\left(1\,,\, p_{M}\tau\left(1-\alpha\right)\right)N_{OPT}+N_{w}
\end{eqnarray}

Inequality (26) is obtained by substituting inequality (23) in equation (25). We now use \textit{Lemma 2} followed by inequality (24) in inequality (27) to get

\vspace{-1.0em}

\begin{eqnarray}
L_{A_{w}} & = & \delta_{M} N_{OPT}+N_{w}\nonumber \\
 & \leq & \delta_{M} N_{OPT}+N_{OPT}\\
 & = & \left(1+\delta_{M}\right)N_{OPT}\\
 & \leq & \left(1+\delta_{M}\right)L_{OPT}
\end{eqnarray}

Therefore the competitive ratio of $A_{w}$ is $c=1+\delta_{M}=1+\min\left(1\,,\, p_{M}\tau\left(1-\alpha\right)\right)$. This proves \textit{Theorem 1}.

\bibliographystyle{IEEEtran}
\bibliography{reference_cloud}

\begin{thebibliography}{10}
\providecommand{\url}[1]{#1}
\csname url@samestyle\endcsname
\providecommand{\newblock}{\relax}
\providecommand{\bibinfo}[2]{#2}
\providecommand{\BIBentrySTDinterwordspacing}{\spaceskip=0pt\relax}
\providecommand{\BIBentryALTinterwordstretchfactor}{4}
\providecommand{\BIBentryALTinterwordspacing}{\spaceskip=\fontdimen2\font plus
\BIBentryALTinterwordstretchfactor\fontdimen3\font minus
  \fontdimen4\font\relax}
\providecommand{\BIBforeignlanguage}[2]{{%
\expandafter\ifx\csname l@#1\endcsname\relax
\typeout{** WARNING: IEEEtran.bst: No hyphenation pattern has been}%
\typeout{** loaded for the language `#1'. Using the pattern for}%
\typeout{** the default language instead.}%
\else
\language=\csname l@#1\endcsname
\fi
#2}}
\providecommand{\BIBdecl}{\relax}
\BIBdecl

\bibitem{berkeley_view}
\BIBentryALTinterwordspacing
M.~Armbrust, A.~Fox, R.~Griffith, A.~D. Joseph, R.~H. Katz, A.~Konwinski,
  G.~Lee, D.~A. Patterson, A.~Rabkin, I.~Stoica, and M.~Zaharia, ``Above the
  clouds: A berkeley view of cloud computing,'' EECS Department, University of
  California, Berkeley, Tech. Rep., 2009. [Online]. Available:
  \url{http://www.eecs.berkeley.edu/Pubs/TechRpts/2009/EECS-2009-28.html}
\BIBentrySTDinterwordspacing

\bibitem{staticthreshold}
M.~Z. Hasan, E.~Magana, A.~Clemm, L.~Tucker, and S.~L.~D. Gudreddi,
  ``Integrated and autonomic cloud resource scaling,'' in \emph{Network
  Operations and Management Symposium (NOMS), 2012 IEEE}.\hskip 1em plus 0.5em
  minus 0.4em\relax IEEE, 2012, pp. 1327--1334.

\bibitem{queue2}
B.~Urgaonkar, P.~Shenoy, A.~Chandra, P.~Goyal, and T.~Wood, ``Agile dynamic
  provisioning of multi-tier internet applications,'' \emph{ACM Transactions on
  Autonomous and Adaptive Systems (TAAS)}, vol.~3, no.~1, p.~1, 2008.

\bibitem{queue3}
A.~Ali-Eldin, J.~Tordsson, and E.~Elmroth, ``An adaptive hybrid elasticity
  controller for cloud infrastructures,'' in \emph{Network Operations and
  Management Symposium (NOMS), 2012 IEEE}.\hskip 1em plus 0.5em minus
  0.4em\relax IEEE, 2012, pp. 204--212.

\bibitem{control3}
X.~Dutreilh, N.~Rivierre, A.~Moreau, J.~Malenfant, and I.~Truck, ``From data
  center resource allocation to control theory and back,'' in \emph{Cloud
  Computing (CLOUD), 2010 IEEE 3rd International Conference on}.\hskip 1em plus
  0.5em minus 0.4em\relax IEEE, 2010, pp. 410--417.

\bibitem{control4}
P.~Padala, K.-Y. Hou, K.~G. Shin, X.~Zhu, M.~Uysal, Z.~Wang, S.~Singhal, and
  A.~Merchant, ``Automated control of multiple virtualized resources,'' in
  \emph{Proceedings of the 4th ACM European conference on Computer
  systems}.\hskip 1em plus 0.5em minus 0.4em\relax ACM, 2009, pp. 13--26.

\bibitem{mpc}
L.~Wang, J.~Xu, M.~Zhao, and J.~Fortes, ``Adaptive virtual resource management
  with fuzzy model predictive control,'' in \emph{Proceedings of the 8th ACM
  international conference on Autonomic computing}.\hskip 1em plus 0.5em minus
  0.4em\relax ACM, 2011, pp. 191--192.

\bibitem{timeseries1}
E.~Caron, F.~Desprez, and A.~Muresan, ``Pattern matching based forecast of
  non-periodic repetitive behavior for cloud clients,'' \emph{Journal of Grid
  Computing}, vol.~9, no.~1, pp. 49--64, 2011.

\bibitem{timeseries2}
Z.~Gong, X.~Gu, and J.~Wilkes, ``Press: Predictive elastic resource scaling for
  cloud systems,'' in \emph{Network and Service Management (CNSM), 2010
  International Conference on}.\hskip 1em plus 0.5em minus 0.4em\relax IEEE,
  2010, pp. 9--16.

\bibitem{PhDThesisML}
P.~Bodik, ``Automating datacenter operations using machine learning,'' Ph.D.
  dissertation, University of California, Berkeley, 2010.

\bibitem{deadline1}
R.~N. Calheiros and R.~Buyya, ``Meeting deadlines of scientific workflows in
  public clouds with tasks replication,'' \emph{Parallel and Distributed
  Systems, IEEE Transactions on}, vol.~25, no.~7, pp. 1787--1796, 2014.

\bibitem{deadline2}
M.~A. Rodriguez and R.~Buyya, ``Deadline based resource provisioningand
  scheduling algorithm for scientific workflows on clouds,'' \emph{Cloud
  Computing, IEEE Transactions on}, vol.~2, no.~2, pp. 222--235, 2014.

\bibitem{queue1}
H.~Khazaei, J.~Misic, and V.~B. Misic, ``Performance analysis of cloud
  computing centers using m/g/m/m+ r queuing systems,'' \emph{Parallel and
  Distributed Systems, IEEE Transactions on}, vol.~23, no.~5, pp. 936--943,
  2012.

\bibitem{control1}
J.~L. Hellerstein, Y.~Diao, S.~Parekh, and D.~M. Tilbury, \emph{Feedback
  control of computing systems}.\hskip 1em plus 0.5em minus 0.4em\relax John
  Wiley \& Sons, 2004.

\bibitem{control2}
T.~F. Abdelzaher, K.~G. Shin, and N.~Bhatti, ``Performance guarantees for web
  server end-systems: A control-theoretical approach,'' \emph{Parallel and
  Distributed Systems, IEEE Transactions on}, vol.~13, no.~1, pp. 80--96, 2002.

\bibitem{control5}
W.~Qin and Q.~Wang, ``Modeling and control design for performance management of
  web servers via an lpv approach,'' \emph{Control Systems Technology, IEEE
  Transactions on}, vol.~15, no.~2, pp. 259--275, 2007.

\bibitem{optimization1}
D.~Ardagna, M.~Trubian, and L.~Zhang, ``Sla based resource allocation policies
  in autonomic environments,'' \emph{Journal of Parallel and Distributed
  Computing}, vol.~67, no.~3, pp. 259--270, 2007.

\bibitem{optimization2}
Z.~Liu, M.~S. Squillante, and J.~L. Wolf, ``On maximizing
  service-level-agreement profits,'' in \emph{Proceedings of the 3rd ACM
  conference on Electronic Commerce}.\hskip 1em plus 0.5em minus 0.4em\relax
  ACM, 2001, pp. 213--223.

\bibitem{electricity}
M.~Polverini, A.~Cianfrani, S.~Ren, and A.~V. Vasilakos, ``Thermal-aware
  scheduling of batch jobs in geographically distributed data centers,''
  \emph{Cloud Computing, IEEE Transactions on}, vol.~2, no.~1, pp. 71--84,
  2014.

\bibitem{cooling}
S.~Ren, ``Batch job scheduling for reducing water footprints in data center,''
  in \emph{Communication, Control, and Computing (Allerton), 2013 51st Annual
  Allerton Conference on}.\hskip 1em plus 0.5em minus 0.4em\relax IEEE, 2013,
  pp. 747--754.

\bibitem{broker}
K.~Vermeersch, ``A broker for cost-efficient qos aware resource allocation in
  ec2,'' Ph.D. dissertation, Master's thesis, University of Antwerp, 2011.

\bibitem{broker_lit}
S.~Nesmachnow, S.~Iturriaga, and B.~Dorronsoro, ``Efficient heuristics for
  profit optimization of virtual cloud brokers,'' \emph{Computational
  Intelligence Magazine, IEEE}, vol.~10, no.~1, pp. 33--43, 2015.

\bibitem{main2}
W.~Wang, D.~Niu, B.~Li, and B.~Liang, ``Dynamic cloud resource reservation via
  cloud brokerage,'' in \emph{Distributed Computing Systems (ICDCS), 2013 IEEE
  33rd International Conference on}.\hskip 1em plus 0.5em minus 0.4em\relax
  IEEE, 2013, pp. 400--409.

\bibitem{main1}
W.~Wang, B.~Li, and B.~Liang, ``To reserve or not to reserve: Optimal online
  multi-instance acquisition in iaas clouds,'' in \emph{Proc. USENIX Intl.
  Conf. Autonomic Computing (ICAC)}, 2013.

\bibitem{main4}
M.~Lin, A.~Wierman, L.~L. Andrew, and E.~Thereska, ``Dynamic right-sizing for
  power-proportional data centers,'' \emph{IEEE/ACM Transactions on
  Networking}, vol.~21, no.~5, pp. 1378--1391, 2013.

\bibitem{MainFuture}
T.~Lu, M.~Chen, and L.~L. Andrew, ``Simple and effective dynamic provisioning
  for power-proportional data centers,'' \emph{Parallel and Distributed
  Systems, IEEE Transactions on}, vol.~24, no.~6, pp. 1161--1171, 2013.

\bibitem{constrained_skirental}
A.~Khanafer, M.~Kodialam, and K.~P.~N. Puttaswamy, ``The constrained ski-rental
  problem and its application to online cloud cost optimization,'' in
  \emph{Proc. IEEE INFOCOM}, 2013, pp. 1492--1500.

\bibitem{main3}
M.~Polverini, S.~Ren, and A.~Cianfrani, ``Capacity provisioning and pricing for
  cloud computing with energy capping,'' in \emph{Proc. Allerton Conference},
  2013, pp. 413--420.

\bibitem{ozdaglar}
I.~Menache, A.~Ozdaglar, and N.~Shimkin, ``Socially optimal pricing of cloud
  computing resources,'' in \emph{VALUETOOLS}, 2011.

\bibitem{algorithmica}
A.~Karlin, C.~Kenyon, and D.~Randall, ``Dynamic tcp acknowledgment and other
  stories about e/(e-1),'' \emph{Algorithmica}, vol.~36, no.~3, pp. 209--224,
  2003.

\bibitem{bahncard}
R.~Fleischer, ``On the bahncard problem,'' \emph{Theoretical Computer Science},
  vol. 268, no.~1, pp. 161--174, 2001.

\bibitem{googlecluster}
\BIBentryALTinterwordspacing
C.~Reiss, J.~Wilkes, and J.~Hellerstein. (2011) Google cluster-usage traces:
  format+schema. [Online]. Available:
  \url{https://code.google.com/p/googleclusterdata/}
\BIBentrySTDinterwordspacing

\bibitem{penalize}
S.~Di and C.-L. Wang, ``Error-tolerant resource allocation and payment
  minimization for cloud system,'' \emph{Parallel and Distributed Systems, IEEE
  Transactions on}, vol.~24, no.~6, pp. 1097--1106, 2013.

\end{thebibliography}

\end{document}